\documentclass[10pt,twocolumn,letterpaper]{article}

\usepackage[pagenumbers]{cvpr} 

\usepackage{times}
\usepackage{epsfig}

\usepackage{threeparttable}
\usepackage{amsmath}
\usepackage{amssymb}
\usepackage{bm}
\usepackage{booktabs}
\usepackage{multirow}
\usepackage{makecell}
\usepackage[accsupp]{axessibility}
\usepackage{cite}
\usepackage{color}

\makeatletter
\DeclareRobustCommand\onedot{\futurelet\@let@token\@onedot}
\def\@onedot{\ifx\@let@token.\else.\null\fi\xspace}

\def\eg{\emph{e.g}\onedot} 
\def\ie{\emph{i.e}\onedot} 
 
\def\etc{\emph{etc}\onedot} 
 
\def\etal{\emph{et al}\onedot}

\makeatother

\newcommand{\fref}[1]{Fig.~\ref{#1}}
\newcommand{\tref}[1]{Table~\ref{#1}}

\usepackage[pagebackref=true,breaklinks=true,colorlinks,bookmarks=false]{hyperref}

\usepackage[capitalize]{cleveref}
\crefname{section}{Sec.}{Secs.}
\Crefname{section}{Section}{Sections}
\Crefname{table}{Table}{Tables}
\crefname{table}{Tab.}{Tabs.}

\def\cvprPaperID{3129} 
\def\confName{CVPR}
\def\confYear{2022}

\begin{document}

\title{Estimating Fine-Grained Noise Model via Contrastive Learning}

\author{Yunhao Zou \qquad \qquad  Ying Fu\thanks{Corresponding Author: fuying@bit.edu.cn}\\
School of Computer Science and Technology, Beijing Institute of Technology\\
}
\maketitle

\begin{abstract}
Image denoising has achieved unprecedented progress as great efforts have been made to exploit effective deep denoisers. To improve the denoising performance in real-world, two typical solutions are used in recent trends: devising better noise models for the synthesis of more realistic training data, and estimating noise level function to guide  non-blind denoisers. In this work, we combine both noise modeling and estimation, and propose an innovative noise model estimation and noise synthesis pipeline for realistic noisy image generation. Specifically, our model learns a noise estimation model with fine-grained statistical noise model in a contrastive manner. Then, we use the estimated noise parameters to model camera-specific noise distribution, and  synthesize realistic noisy training data. The most striking thing for our work is that by calibrating noise models of several sensors, our model can be extended to predict other cameras. In other words, we can estimate camera-specific noise models for unknown sensors with only testing images, without laborious calibration frames or paired noisy/clean data. The proposed pipeline endows deep denoisers with competitive performances with state-of-the-art real noise modeling methods.
\end{abstract}

\section{Introduction}
\label{sec:intro}

Image denoising is a fundamental and significant problem in the community of low-level vision. Taking the advantage of powerful deep learning tools, previous works~\cite{zhang2017beyond, zhang2018ffdnet, tai2017memnet} have achieved nearly perfect performances removing noise under Additive White Gaussian Noise (AWGN) assumption. However, the denoising results on real photographs from consumer-level cameras and mobile devices are less satisfying~\cite{plotz2017benchmarking, abdelhamed2018high, chen2018learning}. This phenomenon is mainly due to the distribution discrepancy between the noise assumption and real sensor noise distribution, which brings large domain gap between training and testing data. To this end, more researchers are dedicated to real noise removal~\cite{guo2019toward, abdelhamed2019noise, brooks2019unprocessing, zhou2020awgn, chen2018image, zeng2021mathrm}. 
There are mainly two significant issues to be solved for real image denoising.

A straightforward way is to model real sensor noise distributions and generate more realistic data~\cite{henz2020synthesizing, yue2020dual, chen2018image, chang2020learning, abdelhamed2019noise, wei2021physics, zhang2021rethinking}. Some methods present statistical models to mimic real noise formation,  they generally calibrate camera-specific noise parameters (\eg, noise variance) from specially captured frames and then generate training data. In this way, deep networks benefit from more realistic training data. Statistical noise models, including AWGN, Poisson-Guassian (P-G, \cite{foi2007noise}) model, Poisson Mixture model~\cite{zhang2017improved}, \etc, are commonly used in the early exploration of noise models. Recently, some noise modeling literatures based on deep generative models like GAN~\cite{henz2020synthesizing, yue2020dual, chen2018image, chang2020learning} and Normalizing Flow~\cite{abdelhamed2019noise} have emerged, but fail in the competition with more fine-grained statistical noise model~\cite{wei2021physics, zhang2021rethinking} with carefully calibrated noise parameters. A limitation for noise modeling methods is that they depend on real calibration frames or noisy/clean pairs of certain camera, which is laborious or unreachable in some scenarios. 

Another important issue is noise estimation. Noise level functions are usually served as guidance for both filter based denoising approaches~\cite{dabov2007image, buades2005non} and deep learning based denoising networks~\cite{zhang2018ffdnet}. Recently, there are several attempts to estimate noise level functions, based on both computation~\cite{chen2015efficient, liu2014practical, liu2006noise, liu2013single, pyatykh2012image, foi2008practical, zhu2016noise, liu2012additive} or deep learning~\cite{byun2021fbi, zhang2018ffdnet, guo2019toward, cao2021pseudo, wu2020unpaired}. Nevertheless, these methods are built upon inferior noise models like AWGN, and cannot be used for the estimation of more complex sensor noise corrupted by circuit readout pattern or source follower. Moreover, existing noise estimation methods basically serve the estimated parameters as an inference input value and feed them into denoising filters~\cite{dabov2007image} or end-to-end deep neural networks~\cite{guo2019toward}. They have not ever tried to exploit more intrinsic attributes of the camera sensor through these parameters. 

In this paper, we propose a novel noise model estimation and noise synthesis pipeline to estimate parameters for fine-grained noise models using only testing data, liberating us from the laborious or unreachable calibration for image sensor.
To achieve this, we present a contrastive noise estimation model to estimate noise parameters from a single image under fine-grained noise model. Our contrastive estimation framework separates each noise component, and well approximates noise parameters of a single image, even if the camera for taking pictures has never been seen by the model.
Then, with the estimated parameters, we are capable to estimate the intrinsic joint distribution of an unknown sensor under state-of-the-art physical noise model. As a result, we apply our pipeline to real image denoising and facilitate the training process by synthesizing more realistic data. Our new camera-specific noise synthesis pipeline relieve the dependencies on sophisticated capturing scheme and generate promising synthetic noisy images. 
The main contributions of our work can be summarized as follows:
\vspace{-1mm}
\begin{enumerate}
	\item We present a novel noise model estimation and realistic noise synthesis pipeline, which can estimate camera noise model only from testing data without any camera-specific training data.
	\vspace{-2mm}
	\item We employ a noise estimation framework based on contrastive learning, which  well approximates parameters for fine-grained noise model.
	\vspace{-2mm}
	\item With our realistic noise synthesis pipeline, deep denoisers can reach competitive results with previous noise generation methods which depend on real noisy/clean images or calibration frames.
\end{enumerate}

\section{Related Work}
\label{sec:related}
In this section, we introduce some works that are most related to the proposed method. First, we review widely used statistical or deep learning based noise modeling methods. Then, we introduce existing approaches and applications of noise estimation.

\vspace{3mm}
\noindent\textbf{Noise Modeling.} 
In recent years, the research of noise removal has been pushed forward greatly via strong deep learning tools. Though denoising under the long-standing AWGN model has been well solved~\cite{buades2005non, aharon2006k, zhang2017beyond}, things go different for denoising images captured by real Digital Single Lens Reflex Camera (DSLR) and sensors of mobile phones. Actually, AWGN is inferior for not taking signal-dependent and complex sensor noise into account. A more precise model is Poisson-Guassian (P-G) model~\cite{foi2007noise}, which considers the unstable photon count on the sensor plane. Heteroscedastic Gaussian (Hetero-G) model~\cite{foi2008practical, guo2019toward} is a widely accepted alternative for P-G, it uses a signal-dependent Gaussian distribution to replace Poisson distribution. Other statistical models including Poisson Mixture model~\cite{zhang2017improved}, mixed AWGN with Random Value Impulse Noise (RVIN) \cite{zhou2020awgn} and Gaussian Mixture Model~\cite{zhu2016noise} are also proposed to model real noise. Recently, Wei~\etal~\cite{wei2021physics} delineate the full picture of sensor noise and craft fine-grained and precise statistical model to describe noise distribution, which greatly boosts the performance in real image denoising, especially in extremely dark imagery. Later, Zhang~\etal~\cite{zhang2021rethinking} directly sample readout signal-independent noise from real bias patches. Deep learning based methods are also presented to implicitly model real sensor noise. For example, generative models like GAN~\cite{goodfellow2014generative} and Normalizing Flow~\cite{Kingma2018glow} have appeared in recent image modeling studies~\cite{abdelhamed2019noise, henz2020synthesizing, chang2020learning, yue2020dual, chen2018image, nam2016holistic, jang2021c2n}. Nevertheless, these methods oversimplify the modern sensor imaging pipeline, and ignore the noise sources corrupted by sensor electronics~\cite{healey1994radiometric, irie2008technique, boie1992analysis}. Moreover, generative models are unstable to train,  and these methods cannot compete with statistical models which are carefully calibrated (\textit{opposite} to directly using the noise parameters recorded in the image profile). These noise modeling methods have special needs of camera-specific data, \eg,  calibration frames or clean/noisy pairs for each target camera. Capturing data and calibrating for each camera sensor can be labor-consuming. In addition, in a multitude of imaging scenarios, these prerequisites are unavailable and cannot be guaranteed.

\vspace{1mm}
\noindent\textbf{Noise Estimation.}  Noise estimation can be used in many denoising methods. For traditional non-blind denoising methods like Non-local Means (NLM)~\cite{buades2005non} and BM3D~\cite{dabov2007image}, noise estimation can be used to predict noise level, which is a required input. In early years, numerous works estimate Gaussian noise level in flat areas~\cite{immerkaer1996fast, meer1990fast}, but they are affected by the size of flat areas. Pyatykh~\etal~\cite{pyatykh2012image} propose a principal component analysis (PCA) based noise level estimation method. Similarly, Chen~\etal~\cite{chen2015efficient} carefully analyze the statistical relationship between noise variance and eigenvalues to estimate Gaussian parameters. In the last decade, some works~\cite{liu2014practical, makitalo2014noise} are proposed to estimate P-G noise from a single image, which is more close to real data. Very recently, Pimpalkhute~\etal~\cite{pimpalkhute2021digital} present a hybrid discrete wavelet transform and edge information removal to estimate Gaussian noise variance. Noise estimation also frequently appears in deep learning based denoising methods~\cite{zhang2018ffdnet, guo2019toward, byun2021fbi}. They typically introduce a noise level estimation module to guide the denoising network with a noise level map. Representative methods, like FFDNet~\cite{zhang2018ffdnet} and CBDNet~\cite{guo2019toward}, use a noise estimation subnetwork consisting of several convolution layers to predict noise map. FBI-Denoiser~\cite{byun2021fbi}
proposed a Poisson-Gaussian Estimation Net to learn the P-G noise parameters solely from noisy images. An inevitable limitation for existing noise estimation methods is that they are built upon less accurate AWGN or P-G noise models. In addition, the estimation of more fine-grained noise models are highly ill-posed, none of these methods can be adapted to estimate such noise model. 

In this work, we aim at estimating noise parameters under a much more complete physics-based noise model, and use those noise parameters for an entirely different purpose. To separate the features of different noise components in the latent space, we devise a data augmentation strategy and learn our estimation model in a contrastive manner~\cite{chen2020simple}. Therefore, we relieve building noise modeling joint distribution from specially captured data.

\begin{figure*}[t]
	\centering
	\includegraphics[width=0.88\linewidth, clip, keepaspectratio
	]{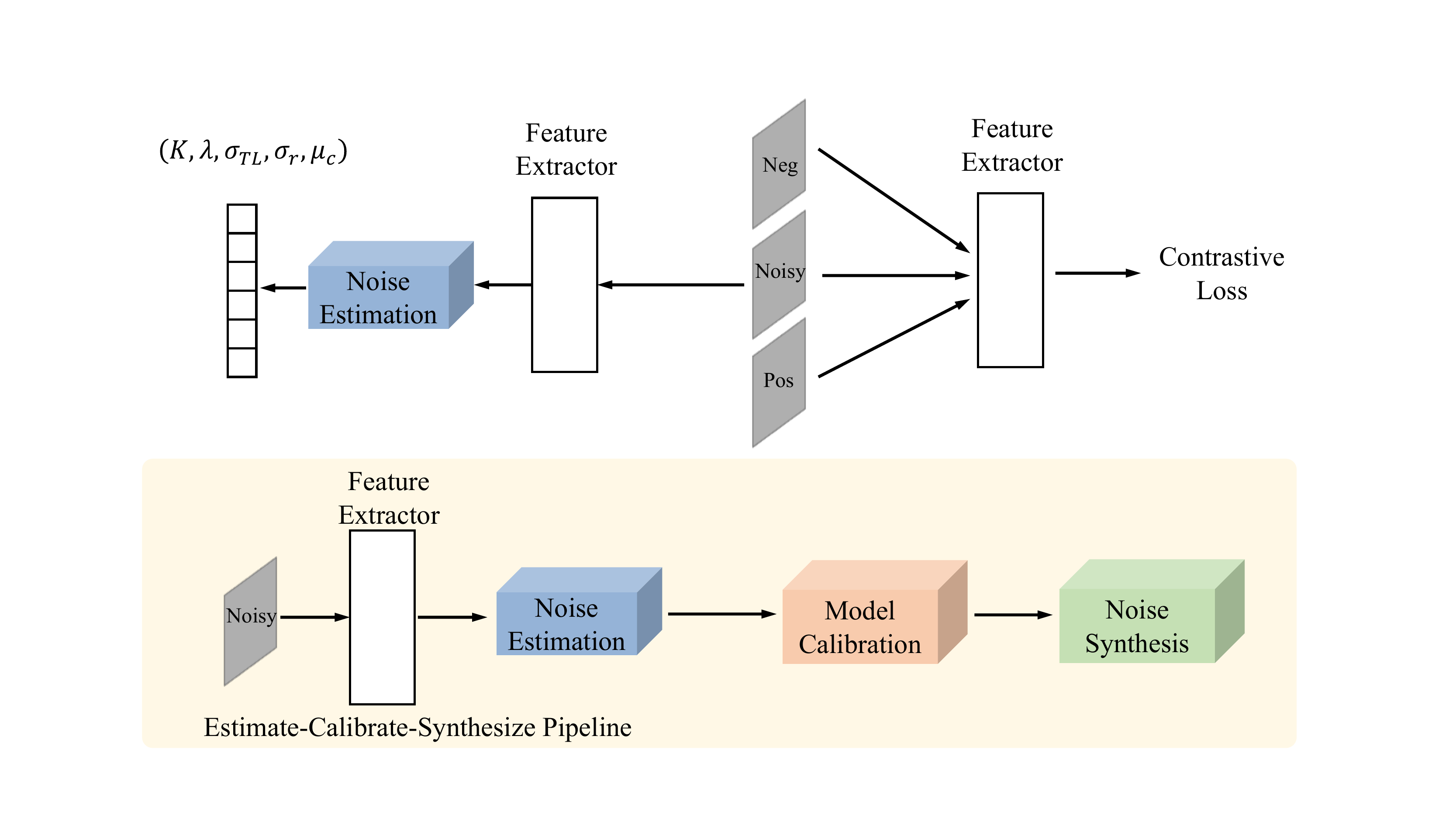}
	\vspace{-2mm}
	\caption{The overview of our camera model estimation and noise synthesis pipeline.}
	\label{fig:overview}
	\vspace{-2mm}
\end{figure*}

\section{Method}
\label{sec:method}
In this section, we first present a fine-grained noise model based on the physical formation of images. Then, we describe our data generation pipeline and contrastive noise estimation framework. The overall pipeline of our work is shown in \fref{fig:overview}.

\subsection{Formulation and Motivation}

Existing real noise generation methods~\cite{wang2020practical, abdelhamed2019noise, chen2018image} suffer from less accurate noise assumption and require laborious calibration frames (\eg, dark frames and flat-field frames) or noisy/clean pairs of a specific camera sensor. In this work, we design a novel noise synthesis pipeline that estimates noise models solely from testing noisy images. Since image noise is mainly produced in linear raw space, in this work, we focus on raw noise modeling and synthesis which is not influenced by image signal processing pipeline (ISP). 

For common CCD and CMOS sensors, the captured raw signal $S$ can be expressed as
\vspace{-1mm}
\begin{equation}
	S=C+N,
	\vspace{-1mm}
\end{equation}
where $C$ and $N$ denote the potential clean image and the summation of all noise components. They are corrupted by the image formation process of CCD/CMOS sensors.

Generally, $N$ has several components, including signal-dependent noise and signal-independent noise, \etc. As a result, $N$ follows a distribution $\mathcal F$ which is decided by the latent clean image $C$
\vspace{-1mm}
\begin{equation}
	N\sim \mathcal F(C).
	\vspace{-1mm}
\end{equation}

The performances of existing data-driven deep learning denoisers are heavily dependent on a large number of $(C, S)$ pairs for supervision. However, the precise formulation of $\mathcal F(C)$ is not reachable, and capturing large real paired dataset is extremely laborious and unbearable. Therefore, many works~\cite{abdelhamed2019noise, henz2020synthesizing, chang2020learning, yue2020dual, chen2018image, nam2016holistic, jang2021c2n} aim at finding a synthetic $\hat N$ which is close to real noise $N$, and accurately modeling the noise distribution $\mathcal F(C)$ is of vital importance.

In this work, we target at solving two significant factors that affect the precision and applicability of existing noise synthesis works, \ie,  less accurate noise models and laborious training data. We also attempt to estimate statistical noise models in situations where cameras can not be reached.

\subsection{Noise Formation Model}
\label{sec:model}
For better noise synthesis, an accurate noise model is indispensable. Here, we formulate a fined-grained noise formation model that is more precise than widely used AWGN and P-G models.

Digital images are corrupted in many steps of electronic imaging pipeline. Among all noise sources, the four most significant components in real-world images are shot noise, readout noise, color bias and row noise~\cite{wei2021physics}.

As is known, due to the quantum nature of light, the number of photons  collected by sensors is unstable. As a result, inevitable shot noise is added to the original photon signal, which follows a Poisson distribution~\cite{hasinoff2014photon}. Given the number of real incident photon $I$, shot noise $N_s$ can be described as
\begin{equation}
	(I+N_s)\sim\mathcal P(I),
\end{equation}
where $\mathcal P$ is the Poisson distribution. Previous works usually replace  $\mathcal P$  with a variance-variant Gaussian distribution, for the purpose of easier calibration. In this work, we are dedicated to real Poisson which is more accurate.

Readout noise is generated when the circuit reads electronic signals and transforms them into voltage level. The combination of different noise sources makes it close to a random Gaussian distribution. In addition, the existence of dark current renders the noise distribution away from zero-centered. On the basis of these considerations, the readout noise $N_{read}$ can be presented as
\vspace{-0.5mm}
\begin{equation}
	N_{read}\sim\mathcal N(\mu_c, \sigma^2),
	\vspace{-0.5mm}
\end{equation}
where $\mu_c$ is the non-zero centered bias. There is obvious color bias in extremely low environment.

Another important component correlated to the electrons-to-voltage process is row noise, which is caused by the row-by-row sensor read out format. 
We model this kind of row noise $N_{row}$ as a Gaussian distribution
\vspace{-0.5mm}
\begin{equation}
	N_{row}\sim\mathcal N(0, \sigma_{r}^2).
	\vspace{-0.5mm}
\end{equation}

Let $K$ denote the overall gain from $I$ to the potential clean image $C$, \ie, $C=KI$, the real-world noise formation model can be expressed as
\vspace{-0.5mm}
\begin{equation}
	N=KN_s+N_{row}+N_{read}.
	\label{eq:noisemodel}
	\vspace{-0.5mm}
\end{equation}

For shot noise, given the overall gain $K$, we can easily obtain noise by reversing digital signal to the number of photons, sampling shot noise from a Poisson distribution, and reversing back to digital signals. Therefore, for the following noise estimation model, we need to estimate a four-tuple noise parameter $(K, \sigma, \mu_c, \sigma_{r})$.

\subsection{Model Estimation and Noise Synthesis Pipeline}
Here, we introduce our innovative model estimation and noise synthesis pipeline. Our pipeline estimates noise model parameters and liberates the noise generation process from the problem of laborious training data and less accurate noise models. Given a testing denoising datasets captured from a single camera sensor, our pipeline first estimates parameters for the noise models mentioned in Section~\ref{sec:model}, and then decides a parameter sampling and noisy image synthesis strategy to generate realistic training data. The whole process rely neither on paired training data nor real calibration frames.

\begin{figure}[t]
	\centering
	
	\begin{subfigure}[b]{.426\linewidth}
		\centering
		\includegraphics[width=\linewidth]{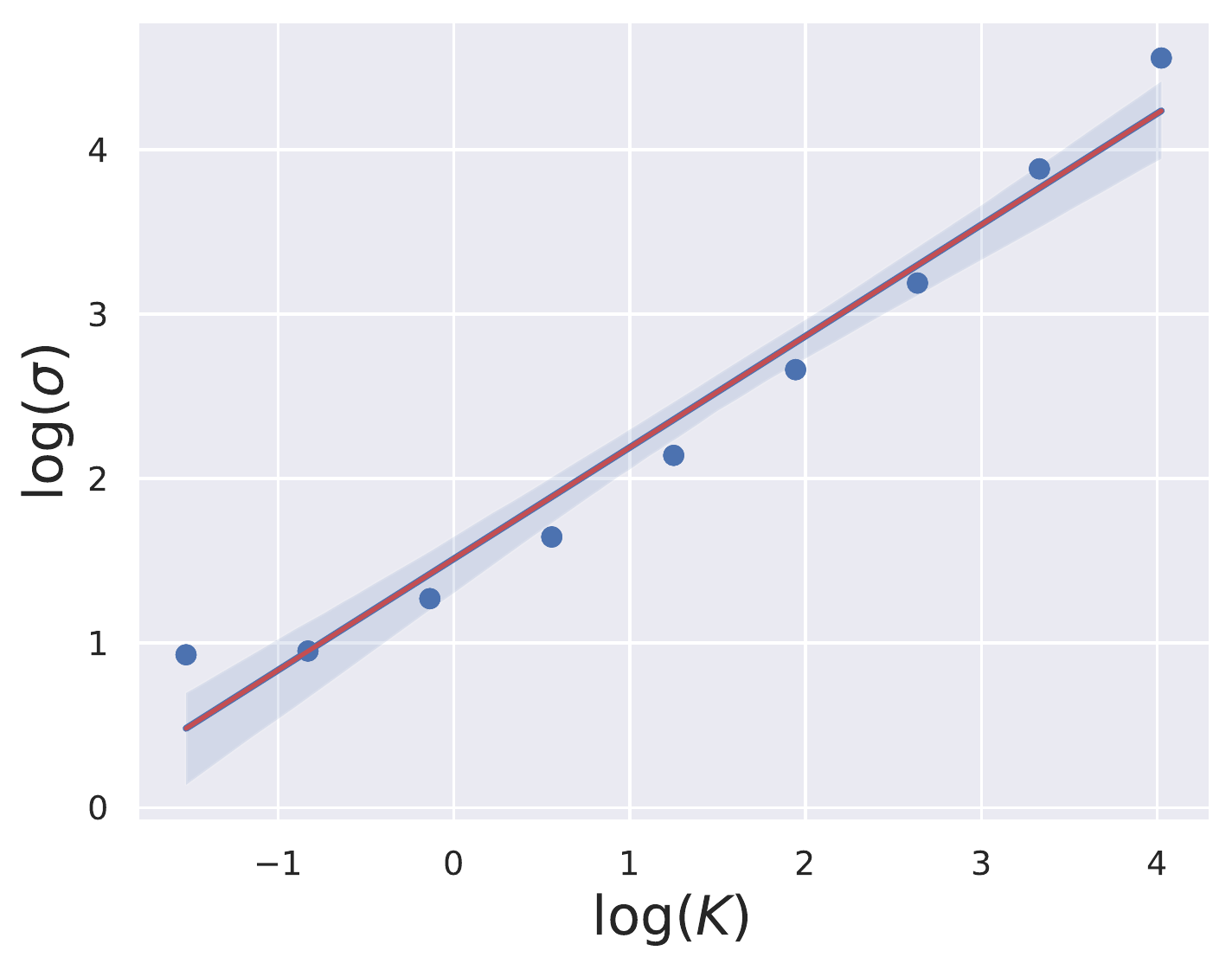}\\
		\caption{Readout Noise}
	\end{subfigure}
	\begin{subfigure}[b]{.45\linewidth}
		\centering
		\includegraphics[width=\linewidth]{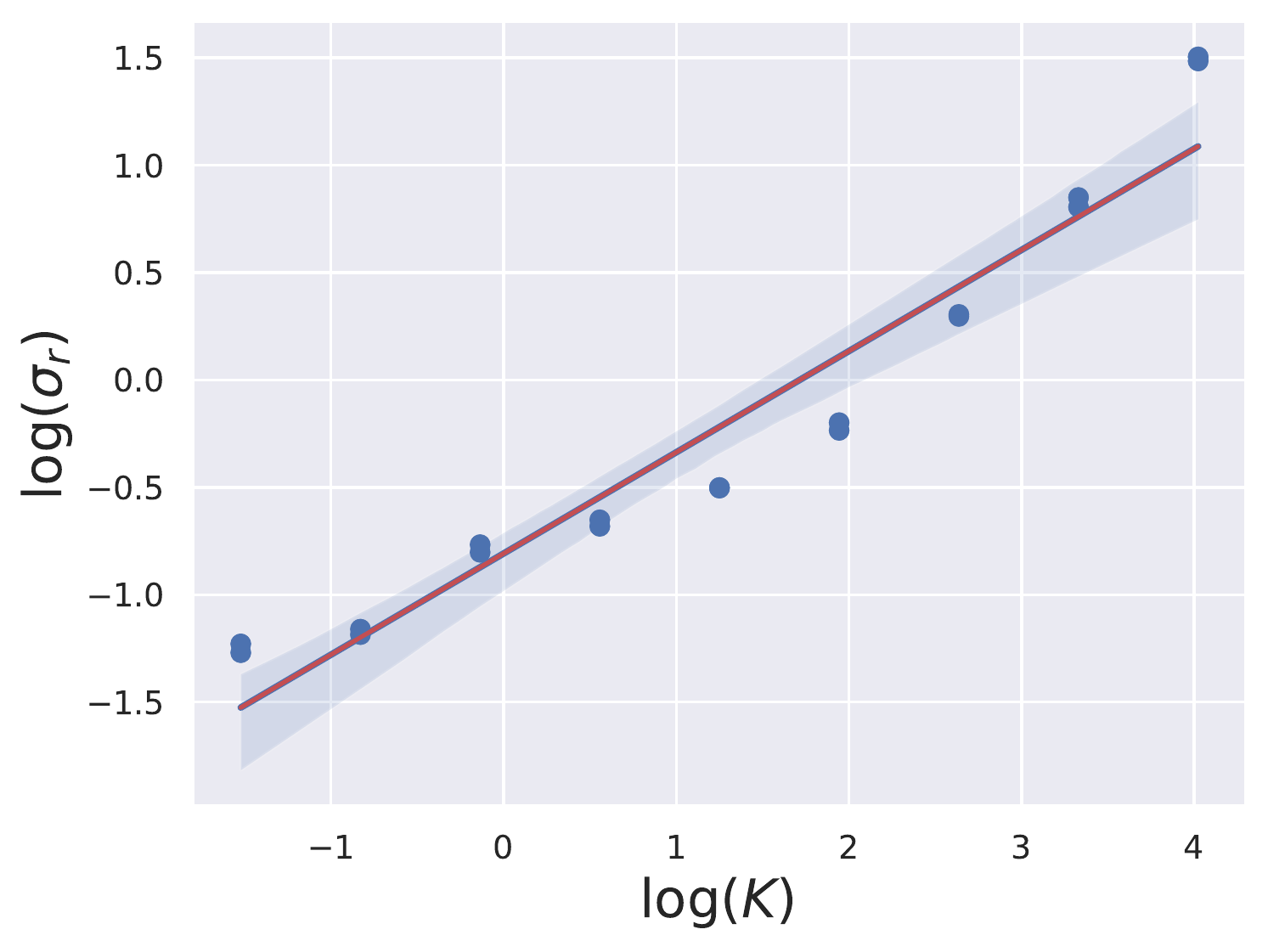}
		\caption{Row Noise}
	\end{subfigure}

	\caption{We provide the calibration results of a typical camera sensor, \ie, Canon EOS5D4, to show the logarithmic linear relationship between $(K, \sigma)$ and $(K, \sigma_r)$.}
	\label{fig:regression}
	\vspace{-3mm}
\end{figure}

The scheme overview of our noise synthesis pipeline is illustrated in Fig.~\ref{fig:overview}. Given a noisy test dataset, we first estimate the noise parameters $\bm P_i=(K, \sigma, \mu_c, \sigma_{r})$ for each single image, and obtain a set of parameter tuples $\{\bm P_1, \bm P_2, \cdots, \bm P_M\}$, where $M$ is the size of testing dataset. Our noise estimation model specially designed for estimating such fine-grained noise would be described in Section~\ref{sec:ne}. According to previous works~\cite{wei2021physics, wang2020practical}, we assume that the system overall gain $K$ is proportional to the ISO setting, and both readout and row noise variance is logarithmically proportional to $K$. Taking Canon EOS5D4 for example, the statistical relationship between noise parameters $\sigma$ and $\sigma_{r}$ can be well fitted by a logarithmic linear model with respect to the overall gain $K$, as shown in \fref{fig:regression}. Therefore, we use a linear regression model to fit the log linear relationship between $K$ and $\sigma$, $\sigma_{r}$, and obtain the estimated bias and slope
\vspace{-1mm}
\begin{equation}
	\begin{aligned}
		&\log\sigma=a\log K + b, \\
		&\log\sigma_{r}=a_{r}\log K + b_{r}. 
	\end{aligned}
\vspace{-1mm}
\end{equation}
From here, we can sample camera gain $K$ from a uniform distribution from the smallest and largest estimated gain in testing sets, denoted as $K_{min}$ and $K_{max}$.  Then, other noise parameters can be sampled following the joint distribution
\vspace{-1mm}
\begin{equation}
	\begin{aligned}
	&\log \left( K \right)  \sim U \left( \log (\hat{K}_{min}), \log (\hat{K}_{max}) \right),   \\   
	&\log \left( \sigma \right) | \log \left( K \right) \sim \mathcal{N} \left(a \log (K)  + b ,  \hat{\sigma}^2 \right), \\  
	&\log \left( \sigma_{r} \right) | \log \left(K \right) \sim \mathcal{N} \left(a_{r} \log (K)  + b_{r} ,  \hat{\sigma}_{r}^2 \right),   
	\end{aligned}
\label{eq: sampling}
\vspace{-1mm}
\end{equation}
where $\hat{\sigma}$ and $\hat{\sigma}_{r}$ are the unbiased estimation for noise standard deviation.

If it is necessary to synthesize realistic training samples under any given ISO value $O$ (not limited to discrete ISO values), we can use a linear model to fit the relationship between $K$ and $O$, \ie, $K = \alpha \cdot O$, and replace the sampling strategy of $K$ in Eq.~\eqref{eq: sampling}.

\subsection{Contrastive Noise Estimation Model}
\label{sec:ne}

Although previous noise estimation methods obtain satisfactory performance under AWGN and P-G noise assumptions, such noise models are coarse noise models. Besides, previous noise estimation methods have not tried estimating a more complex and accurate noise assumption. In this section, we propose a deep learning based noise estimation model to predict the four-tuple noise parameters $(K, \sigma, \mu_c, \sigma_{r})$ from a single noisy image.

Actually, the problem of estimating the noise parameters in Eq.~\eqref{eq:noisemodel} is highly ill-posed and hard to be statistically solved by existing PCA-based~\cite{pyatykh2012image} or decomposition-based~\cite{chen2015efficient} noise estimation methods. Moreover, this problem is challenging even for deep neural networks, for deep networks need to distinguish each noise component and estimate noise level from different dimensions. To tackle this problem, we employ a contrastive learning strategy. We first learn an extractor to extract the most discriminative representation for noise estimation, regardless of low frequency scene information.  By contrasting scenes with the same or different noise parameters, it is easier for noise estimation networks to learn precise parameter values.

We employ a simple and efficient  framework~\cite{chen2020simple} for contrastive learning. It learns feature representations by maximizing agreement between differently augmented views of the same label via a contrastive loss in the projection space.  Our contrastive noise estimation framework is illustrated at the bottom of \fref{fig:overview}. The learning process has two stages, including an unsupervised \textit{contrastive feature learning} stage (bottom right) and a supervised \textit{noise estimation} stage (bottom left). Besides, we need a stochastic \textit{data augmentation} strategy to synthesize positive and negative samples. The main components for our contrastive noise estimation model are described in the following.

\vspace{1mm}
\noindent\textbf{Data augmentation.} Given an anchor noisy image $S_i$ synthesized under the $i$-th scene $C_i$ and parameter $P_i$, the feature extractor needs to be fed with positive and negative data samples. In our case, positive samples share the same noise parameters with the anchor image, while negative samples are synthesized with different noise parameters. In addition, to avoid the influence of scenes, both samples are sampled from a random scene. As a result, positive sample $S_i^+$ and negative sample $S_i^-$ are synthesized under $(C_k, P_i)$ and $(C_j, P_j)$. Considering that the information of noise levels are typically drawn from the frequency components along global, vertical or horizontal dimension, we employ a Haar wavelet transformation $t(\cdot)$ before the feature extractor~\cite{haar1910theorie}.

\vspace{1mm}
\noindent\textbf{Contrastive feature learning.}
A feature extractor $f(\cdot)$ is used to extract representations from the frequency image $t(S)$. For sake of simplicity, we use ResNet as the feature extractor backbone, and obtain feature $\bm h=f(t(S))$ for each sample. After that, a small multi-layer perception (MLP) $g(\cdot)$ is used to project representations to low-dimensional vector, and we obtain $\bm z$, $\bm z^+$ and $\bm z^-$ for the anchor, positive and negative sample, respectively. Then, the contrastive framework learns to enlarge the similarity between $(\bm z, \bm z^+)$, and decrease it between  $(\bm z, \bm z^-)$. The similarity calculation function $s$ can be any distance function, and here we utilize cosine similarity. The loss for contrastive learning can be represented as
\begin{equation}
	\mathcal L_{\text{contrastive}}=-\log\frac{\exp(s(\bm z, \bm z^+)/\tau)}{\sum\exp (s(\bm z, \bm z^\pm)/\tau)},
\end{equation}
where $\tau$ denotes the temperature parameter.

\vspace{1mm}
\noindent\textbf{Noise estimation.}
By minimizing the contrastive loss $\mathcal L_{\text{contrastive}}$, the feature extractor $f$ is able to learn the discriminative noise feature of input noisy images. As for our supervised noise estimation learning, we directly add a prediction tail that consists of fully connected layers on the extracted feature $\bm h$. In the training stage, the contrastive representation learning framework is trained first. Then, the noise estimation module is added and trained together with the encoder. We utilize Mean Squre Error (MSE) loss on the predicted noise parameters. Instead of directly penalizing on the predicted $\hat P_i$, we employ a transformation $r$ to balance the weight and scale of $\hat P_i$. In the experiment, we operate logarithm on $\sigma$ and $\sigma_r$, and set weights to $(1, 1, 10, 10)$ for $(K, \log \sigma, \mu_c, \log \sigma_{r})$. Therefore, the learning loss can be formulated as
\vspace{-1mm}
\begin{equation}
	\mathcal L=\sum_i^M\Vert r(P_i)-r(\hat P_i)\Vert_2^2 +\tau \mathcal L_{\text{contrastive}},
	\vspace{-1mm}
\end{equation}
where $M$ is the number of training samples, and $\tau$ is set to $0.1$ in the experiment.

\section{Experiments}
\label{sec:exp}

In this section, we first provide the experimental settings, including the used evaluation metrics and datasets. Then, we conduct experiments on our noise estimation and synthesis pipeline, as well as the downstream denoising task. Finally, we conduct experiments for ablation study.
\subsection{Experimental Setting}

\noindent\textbf{Metrics.}
For noisy image synthesis, we use KL divergence to evaluate the distance between synthetic noise and noisy data captured by real camera sensor. We follow previous work~\cite{abdelhamed2019noise} to perform discrete KL divergence between the histogram of noise patches, which can be formulated as $\sum p(x_i)\log(p(x_i)/q(x_i))$, where  $p(x_i)$ and $q(x_i)$ are the normalized histogram bins of real and estimated samples. As for real denoising experiments, we utilize Signal-to-Noise Ratio (PSNR) and Structural Similarity (SSIM), which are used to measure the 2D spatial fidelity. Larger PSNR and SSIM suggest better results, while smaller KL divergence shows better synthesis.

\vspace{1mm}
\noindent\textbf{Dataset.} Our pipeline is evaluated on a widely used real image denoising dataset SIDD~\cite{abdelhamed2018high}.  SIDD is collected by five smartphone cameras, including Samsung Galaxy S6 Edge (S6), iPhone 7 (IP), Google Pixel (GP), Motorola Nexus 6 (N6) and LG G4 (G4). It contains 320 RAW image pairs for training and testing. In addition, we also synthesize noise on other public  paired raw datasets, including CRVD~\cite{yue2020supervised} and PMRID~\cite{wang2020practical}, aiming to prove the generalization of our noise synthesis pipeline. To train our noise estimation network, we follow calibration steps~\cite{wei2021physics} to carefully calibrate several camera sensors through real bias and flat-field frames, which make up our camera noise model dataset. Specifically, as the noise components in Eq.~\eqref{eq:noisemodel} are additive, we calibrate them one by one. After the calibration of a former component, the mean value of this noise is subtracted, to avoid affecting the calibration of other noise components. Our camera noise model dataset includes Canon EOS 5D4, Nikon D850, Sony RX100VI and HUAWEI P40 Pro.

\vspace{1mm}
\noindent\textbf{Implementations.} In the experiment, the losses are minimized with the adaptive moment estimation method~\cite{kingma2014adam}, with the  momentum parameter 0.9. The learning rate is initially set to $10^{-4}$ , and divided by 10 every 50 epochs. Since larger batch size benefits the learning of contrastive framework, we set batch size as $32$ in the training stage. Both estimation and denoising process are trained for 200 epochs. Our model is implemented using the deep learning framework PyTorch~\cite{paszke2019pytorch}, and we use an NVIDIA RTX 3090 GPU to train our model.

\subsection{Noise Model Estimation and Noise Synthesis}
\label{sec:estimation}

\begin{figure}[t]
	\centering
	
	\begin{subfigure}[b]{.43\linewidth}
		\centering
		\includegraphics[width=\linewidth]{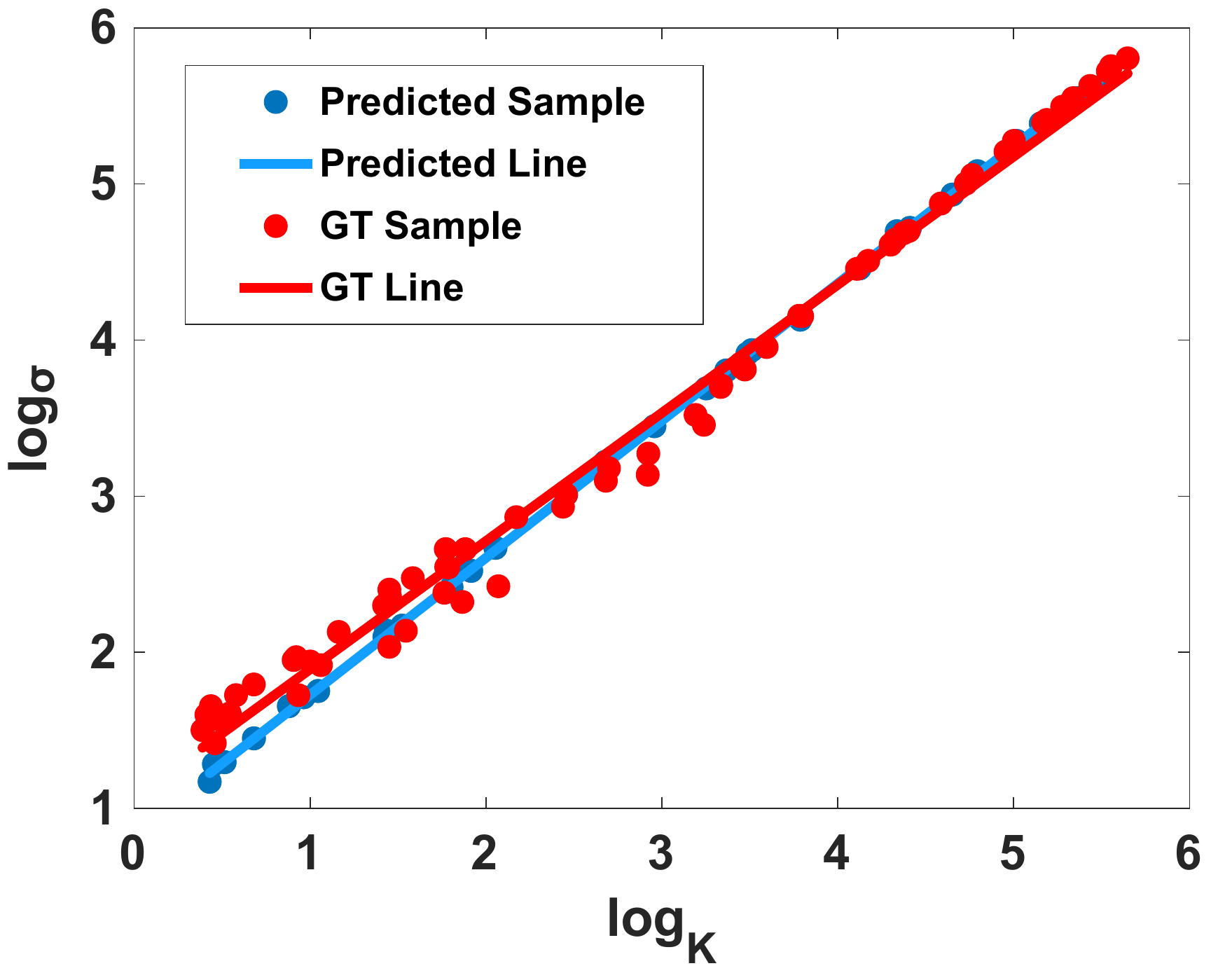}
		\caption{Sony RX100VI}
	\end{subfigure}
	\begin{subfigure}[b]{.45\linewidth}
		\centering
		\includegraphics[width=\linewidth]{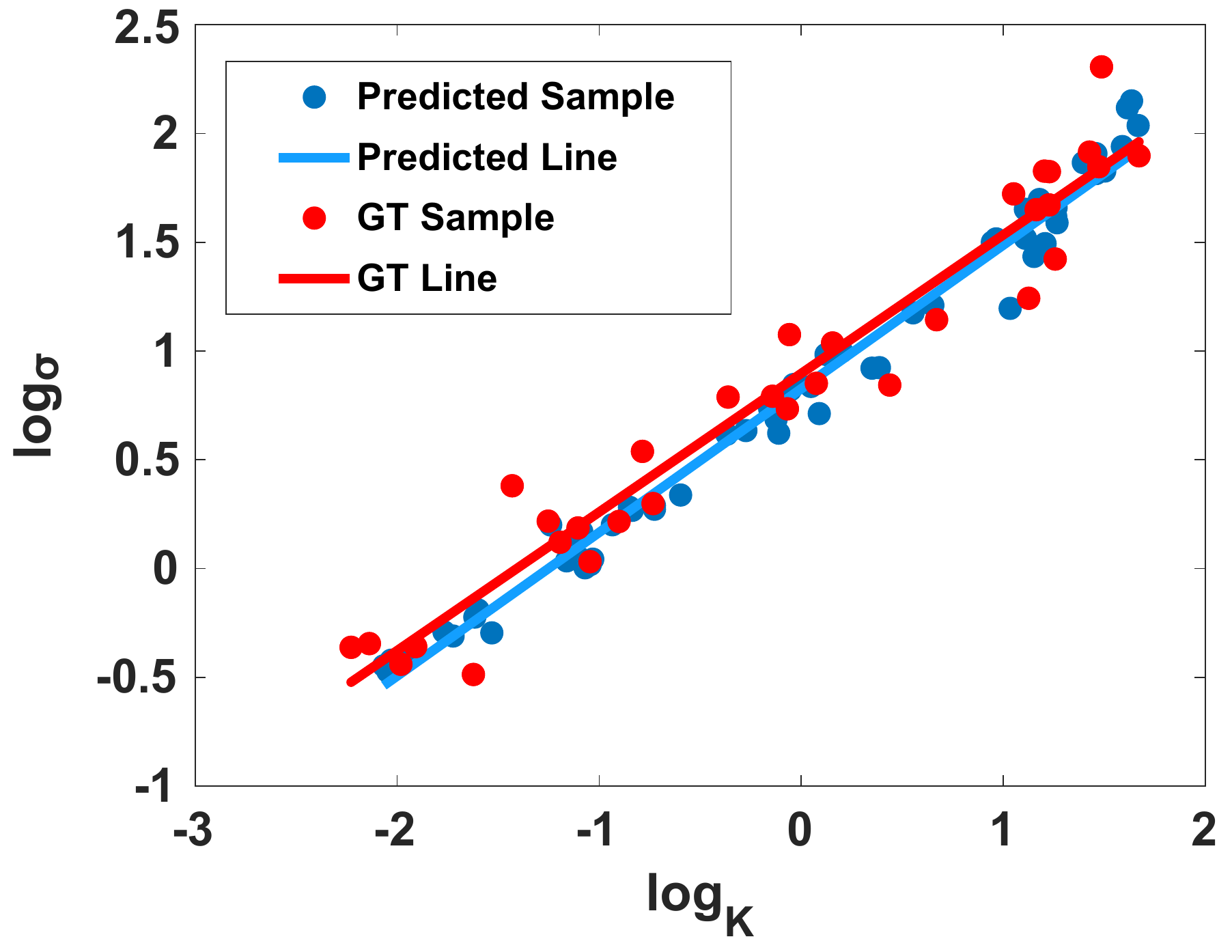}
		\caption{HUAWEI P40 Pro}
	\end{subfigure}
	\begin{subfigure}[b]{.44\linewidth}
		\centering
		\includegraphics[width=\linewidth]{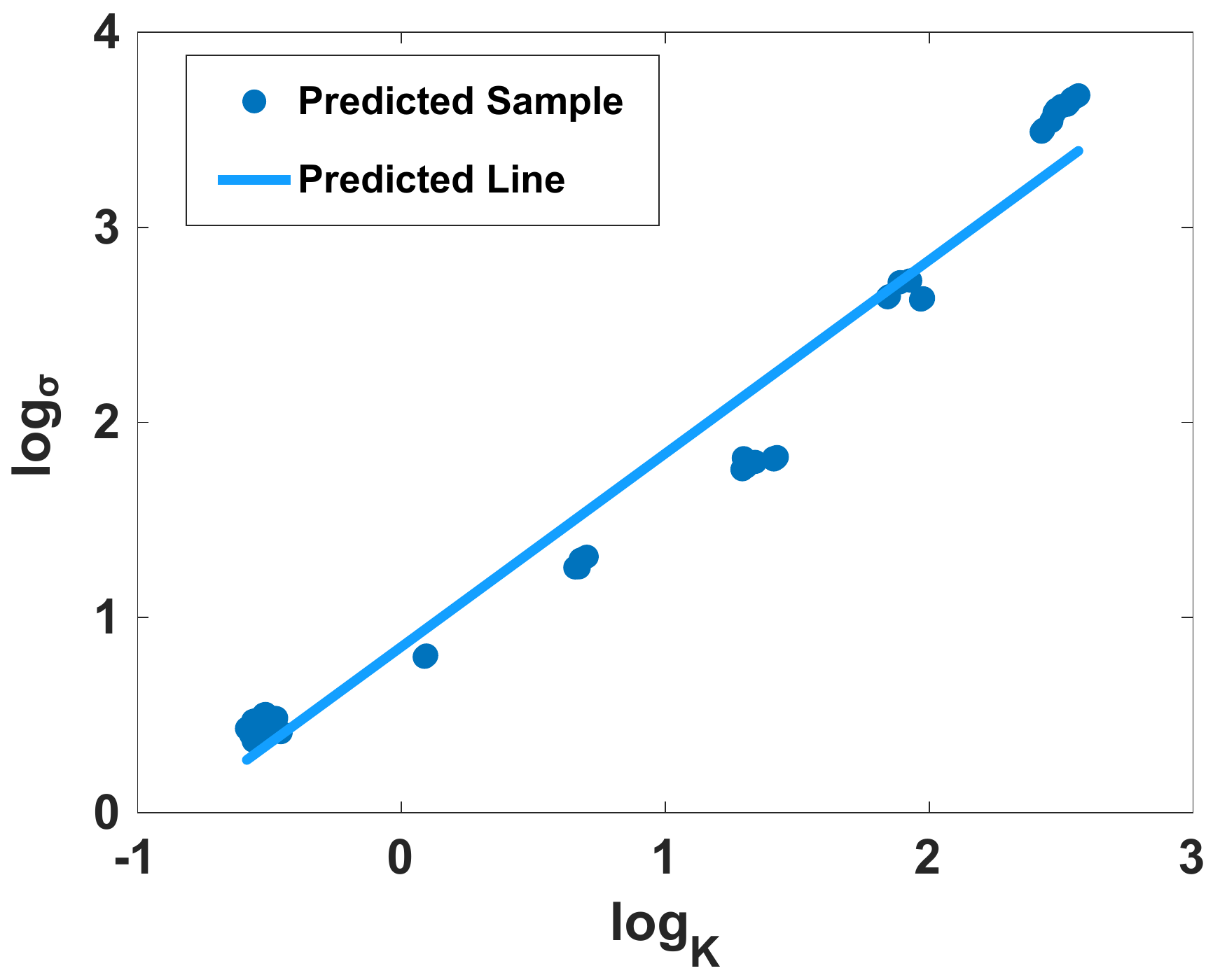}
		\caption{Samsung Galaxy S6}
	\end{subfigure}
	\begin{subfigure}[b]{.455\linewidth}
		\centering
		\includegraphics[width=\linewidth]{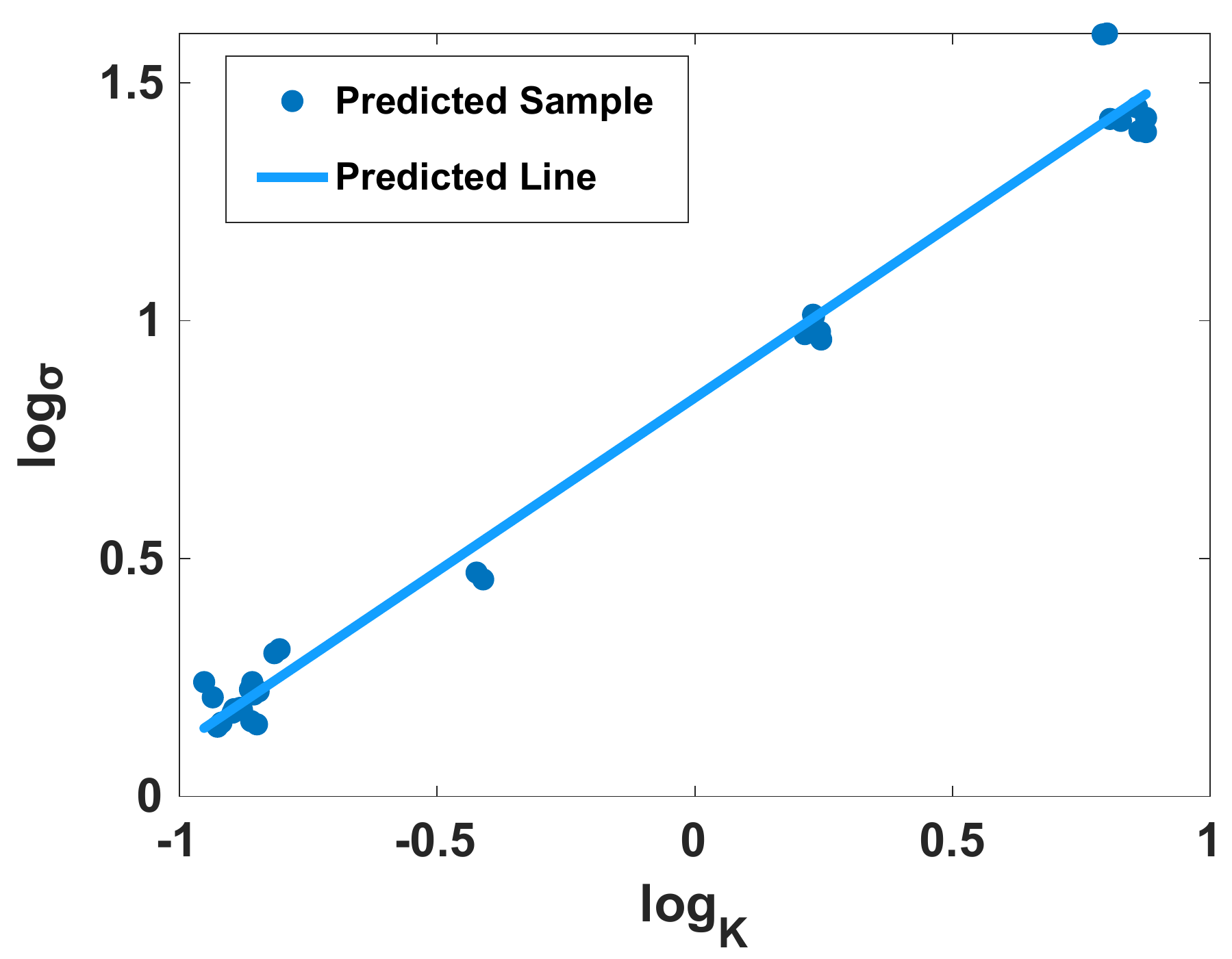}
		\caption{LG G4}
	\end{subfigure}
	\vspace{-1mm}
	\caption{Noise model estimation Performances. The top two camera models are estimated on synthetic noisy images, while the bottom camera models are estimated on real SIDD datasets where no Ground Truth parameters are available.}
	\label{fig:estimation_plot}
	\vspace{-4mm}
\end{figure}

\begin{table*}[t]
	\centering
	\setlength{\tabcolsep}{3.5mm}
	\small
	
	\begin{threeparttable}
		\vspace{-2mm}
		\caption{The performance of noise synthesis for all compared methods on SIDD datasets. The quantitative results for five SIDD cameras are evaluated in KL divergence. Our method provides marginal improvements over other noise synthesis methods, without feeding any camera-specific training data. The best results are highlighted in \textbf{bold}.}
		\vspace{-4mm}
		\begin{tabular}{p{1cm}<{\centering}p{2cm}<{\centering}p{2cm}<{\centering}p{2cm}<{\centering}p{2cm}<{\centering}p{2cm}<{\centering}}
			\toprule
			 Camera&AWGN&P-G&Noiseflow~\cite{abdelhamed2019noise}&CANGAN~\cite{chang2020learning}&Ours\\
			\midrule
			S6 & $0.4793$ & $0.1023$ & $0.0617$ & $0.0432$ & $\bm{0.0385}$\\
			IP & $0.8367$ & $0.0514$ & $0.0327$ & $0.0178$ & $\bm{0.0100}$\\
			GP & $0.6254$ & $0.0316$ & $0.0756$ & $\bm{0.0146}$ & $0.0219$\\
			N6 & $0.7321$ & $0.0168$ & $0.0731$ & $0.0187$ & $\bm{0.0165}$\\
			G4 & $1.0987$ & $0.0315$ & $0.0519$ & $\bm{0.0161}$ & $0.0187$\\
			\midrule
			Average & $0.7544$	& $0.0467$	& $0.0590$	& $0.0220$	& $\bm{0.0211}$\\
			\bottomrule
		\end{tabular}
		\label{tab:synthesis}
	\end{threeparttable}
\end{table*}

\begin{figure*}[h!] \small
	\centering
	\setlength{\tabcolsep}{1pt}
	\begin{tabular}{ccccccc}
		AWGN & P-G & Noiseflow~\cite{abdelhamed2019noise} & CANGAN~\cite{chang2020learning} & Ours & Real & Clean\\
		\multicolumn{1}{m{.125\linewidth}}{\includegraphics[width=\linewidth,clip,keepaspectratio]{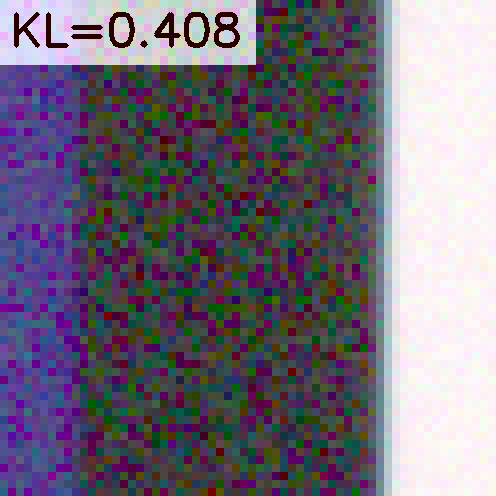}} &
		\multicolumn{1}{m{.125\linewidth}}{\includegraphics[width=\linewidth,clip,keepaspectratio]{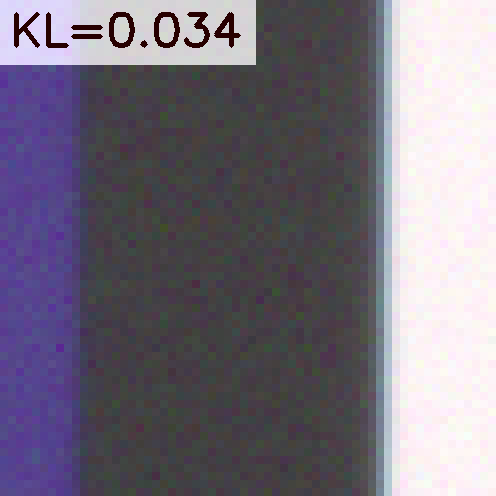}} & \multicolumn{1}{m{.125\linewidth}}{\includegraphics[width=\linewidth,clip,keepaspectratio]{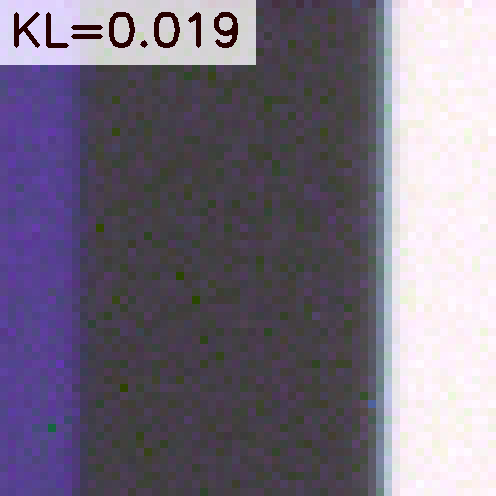}} & \multicolumn{1}{m{.125\linewidth}}{\includegraphics[width=\linewidth,clip,keepaspectratio]{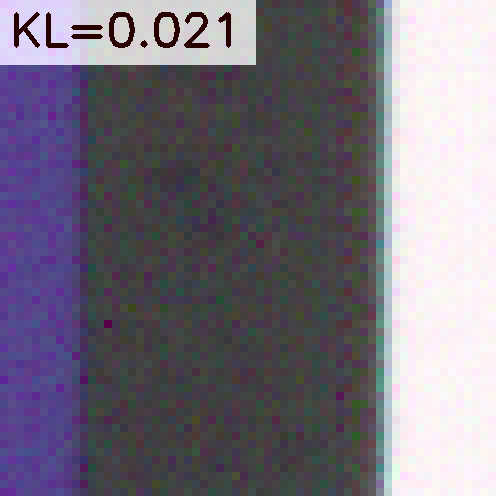}} & \multicolumn{1}{m{.125\linewidth}}{\includegraphics[width=\linewidth,clip,keepaspectratio]{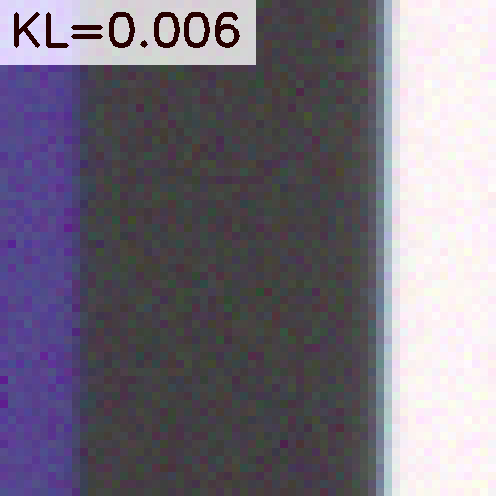}} & \multicolumn{1}{m{.125\linewidth}}{\includegraphics[width=\linewidth,clip,keepaspectratio]{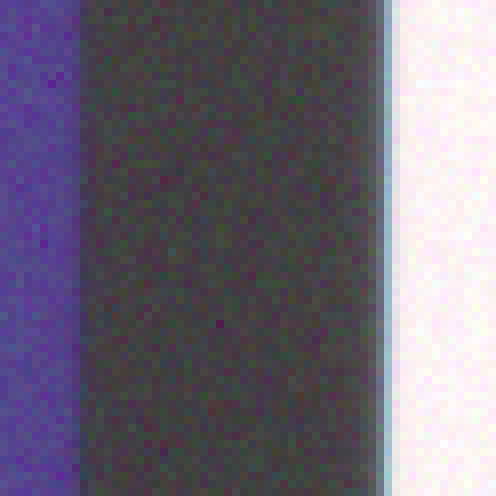}} & \multicolumn{1}{m{.125\linewidth}}{\includegraphics[width=\linewidth,clip,keepaspectratio]{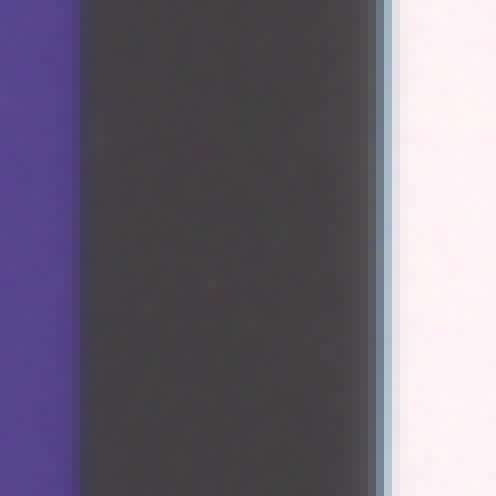}}\\ 
		\multicolumn{1}{m{.125\linewidth}}{\includegraphics[width=\linewidth,clip,keepaspectratio]{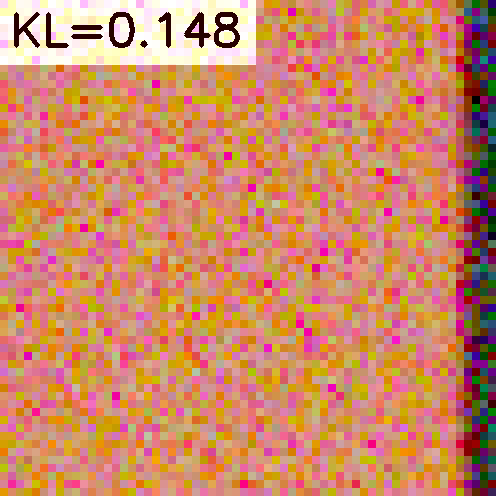}} &
		\multicolumn{1}{m{.125\linewidth}}{\includegraphics[width=\linewidth,clip,keepaspectratio]{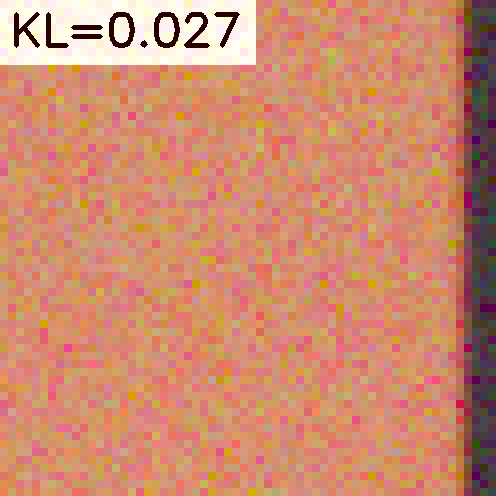}} & \multicolumn{1}{m{.125\linewidth}}{\includegraphics[width=\linewidth,clip,keepaspectratio]{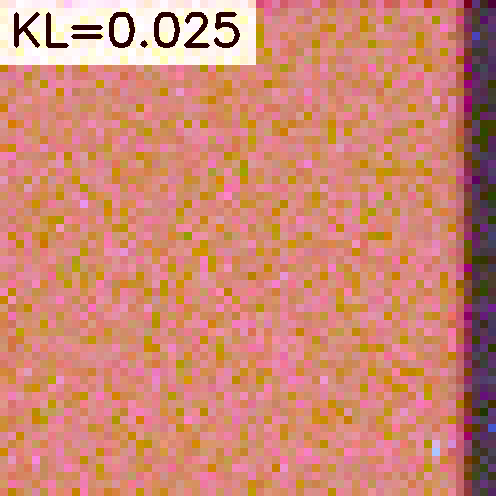}} & \multicolumn{1}{m{.125\linewidth}}{\includegraphics[width=\linewidth,clip,keepaspectratio]{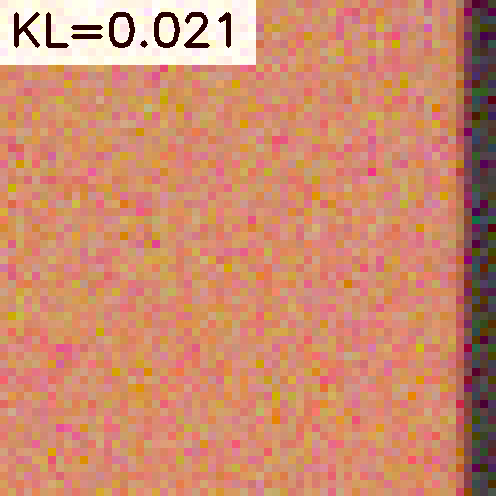}} & \multicolumn{1}{m{.125\linewidth}}{\includegraphics[width=\linewidth,clip,keepaspectratio]{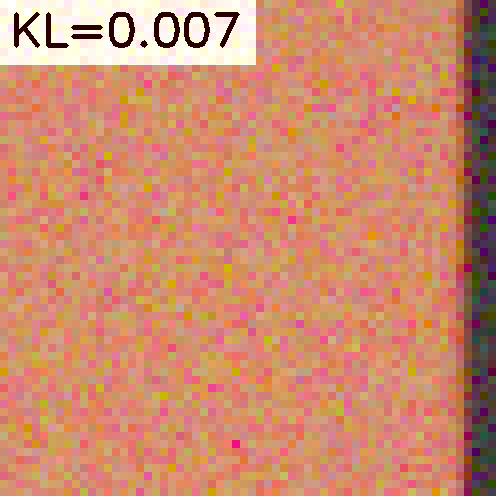}} & \multicolumn{1}{m{.125\linewidth}}{\includegraphics[width=\linewidth,clip,keepaspectratio]{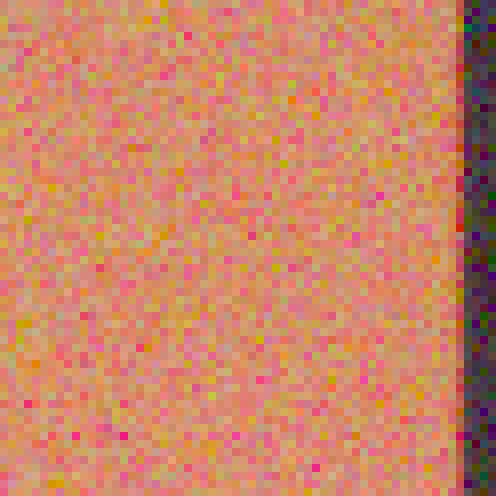}} & \multicolumn{1}{m{.125\linewidth}}{\includegraphics[width=\linewidth,clip,keepaspectratio]{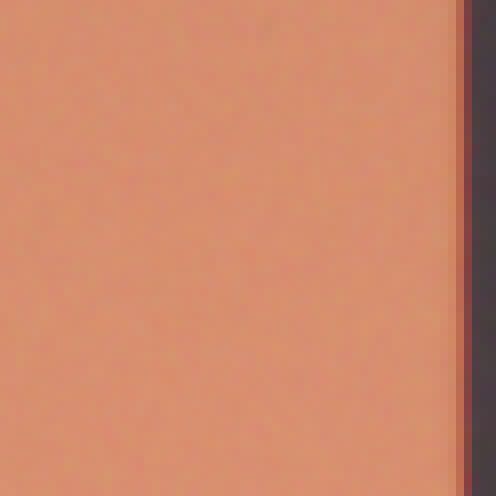}}\\ 
		\multicolumn{1}{m{.125\linewidth}}{\includegraphics[width=\linewidth,clip,keepaspectratio]{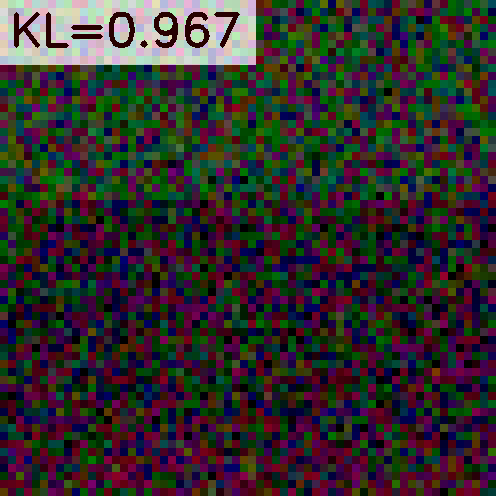}} &
		\multicolumn{1}{m{.125\linewidth}}{\includegraphics[width=\linewidth,clip,keepaspectratio]{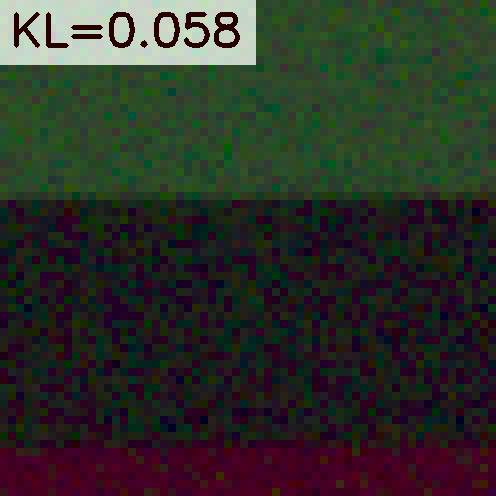}} & \multicolumn{1}{m{.125\linewidth}}{\includegraphics[width=\linewidth,clip,keepaspectratio]{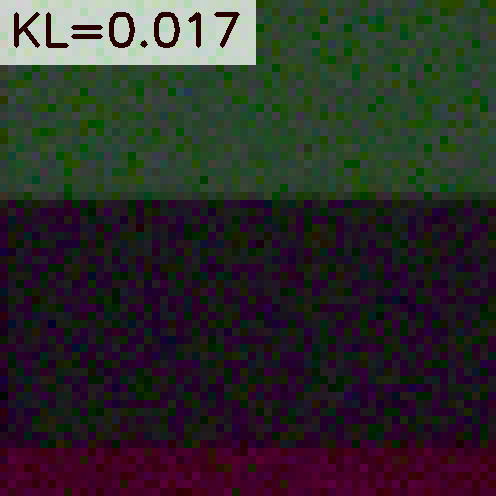}} & \multicolumn{1}{m{.125\linewidth}}{\includegraphics[width=\linewidth,clip,keepaspectratio]{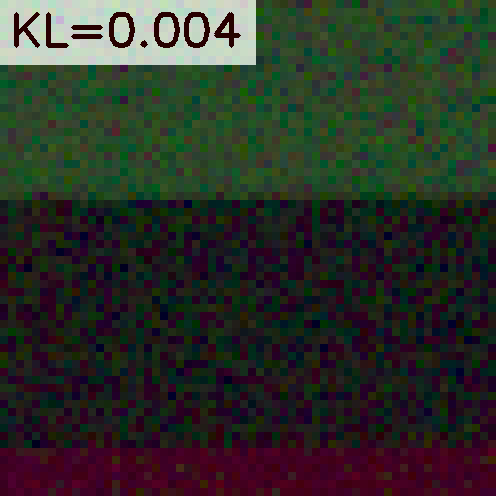}} & \multicolumn{1}{m{.125\linewidth}}{\includegraphics[width=\linewidth,clip,keepaspectratio]{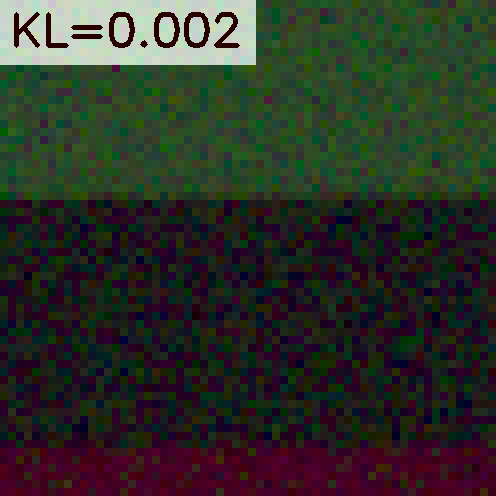}} & \multicolumn{1}{m{.125\linewidth}}{\includegraphics[width=\linewidth,clip,keepaspectratio]{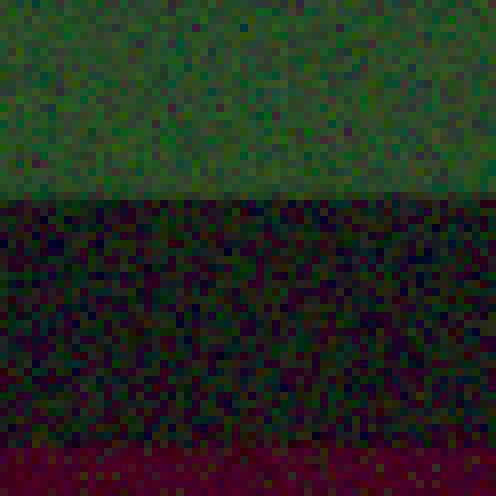}} & \multicolumn{1}{m{.125\linewidth}}{\includegraphics[width=\linewidth,clip,keepaspectratio]{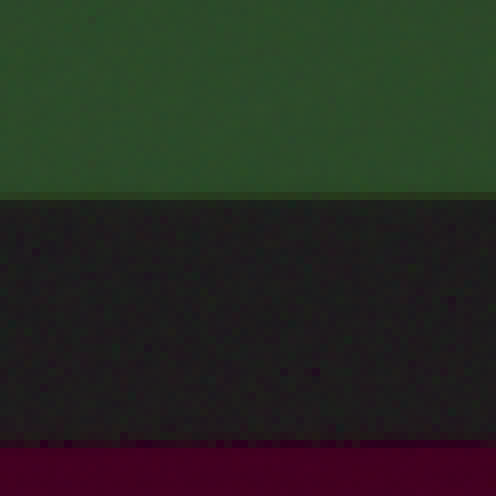}}\\ 
		\multicolumn{1}{m{.125\linewidth}}{\includegraphics[width=\linewidth,clip,keepaspectratio]{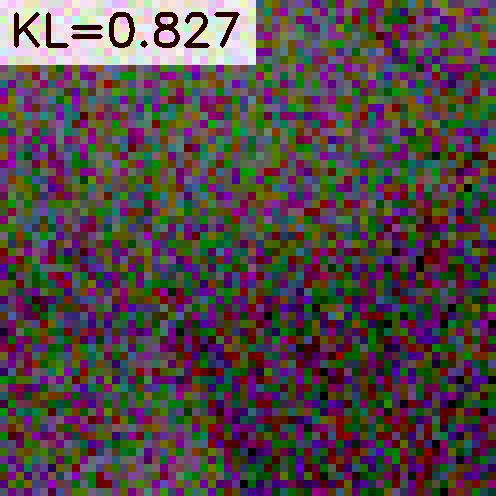}} &
		\multicolumn{1}{m{.125\linewidth}}{\includegraphics[width=\linewidth,clip,keepaspectratio]{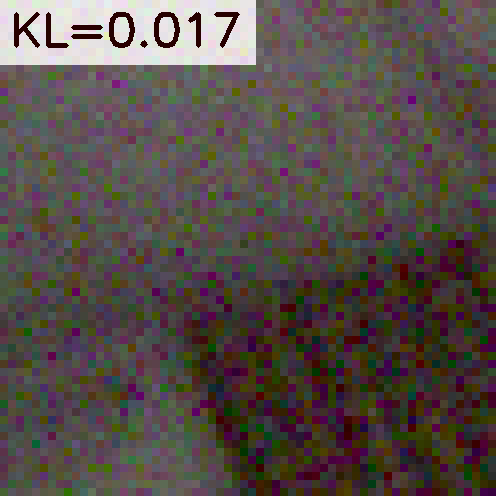}} & \multicolumn{1}{m{.125\linewidth}}{\includegraphics[width=\linewidth,clip,keepaspectratio]{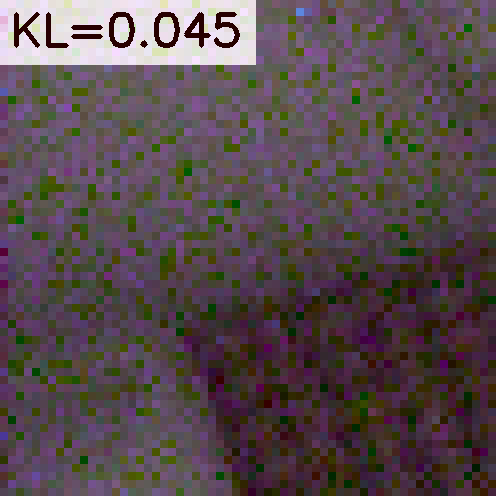}} & \multicolumn{1}{m{.125\linewidth}}{\includegraphics[width=\linewidth,clip,keepaspectratio]{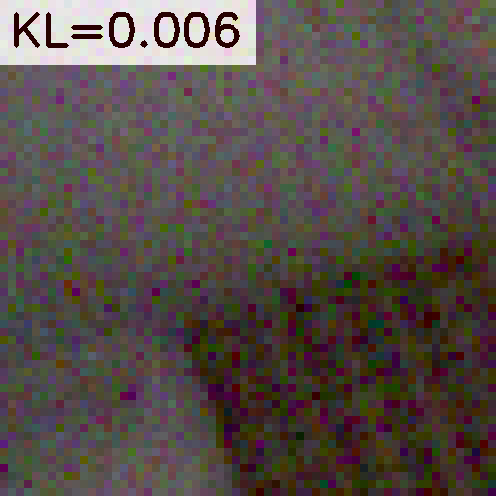}} & \multicolumn{1}{m{.125\linewidth}}{\includegraphics[width=\linewidth,clip,keepaspectratio]{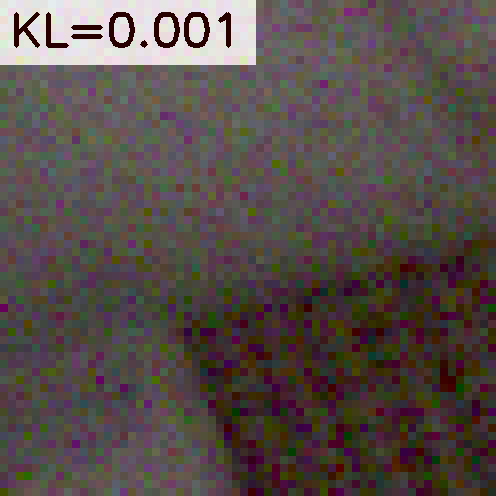}} & \multicolumn{1}{m{.125\linewidth}}{\includegraphics[width=\linewidth,clip,keepaspectratio]{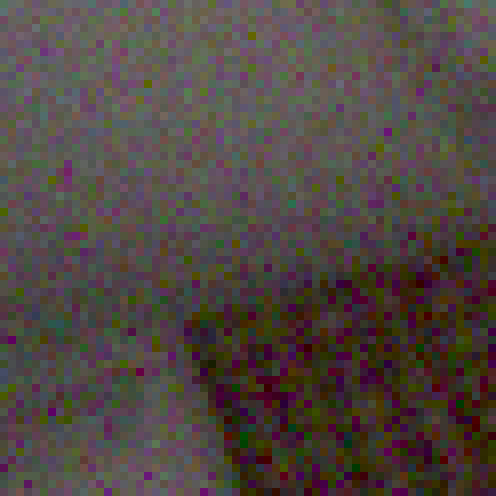}} & \multicolumn{1}{m{.125\linewidth}}{\includegraphics[width=\linewidth,clip,keepaspectratio]{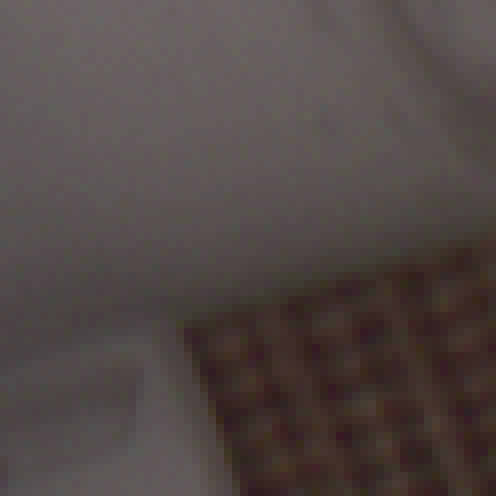}}\\

	\end{tabular}
	\vspace{-2mm}
	\caption{The synthesized noisy images on SIDD dataset~\cite{abdelhamed2018high}. The results of AWGN/P-G/Noiseflow/CANGAN/ Ours/Real Noisy image/Clean input images are shown from left to right.}
	\label{fig:synthesis}
\end{figure*}

\begin{figure}[t] \small
	\centering
	\setlength{\tabcolsep}{1pt}
	\begin{tabular}{cccc}
		\rotatebox{90}{\qquad \quad CRVD} & \includegraphics[width=.29\linewidth,clip,keepaspectratio]{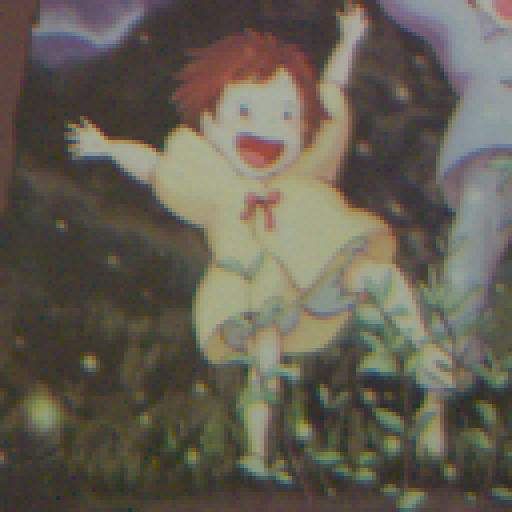} & 
		\includegraphics[width=.29\linewidth,clip,keepaspectratio]{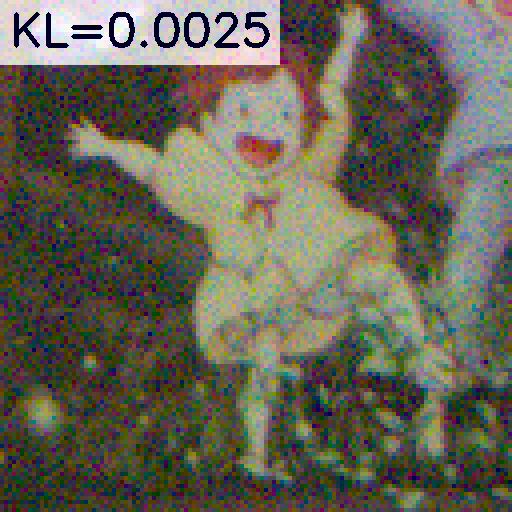} & 
		\includegraphics[width=.29\linewidth,clip,keepaspectratio]{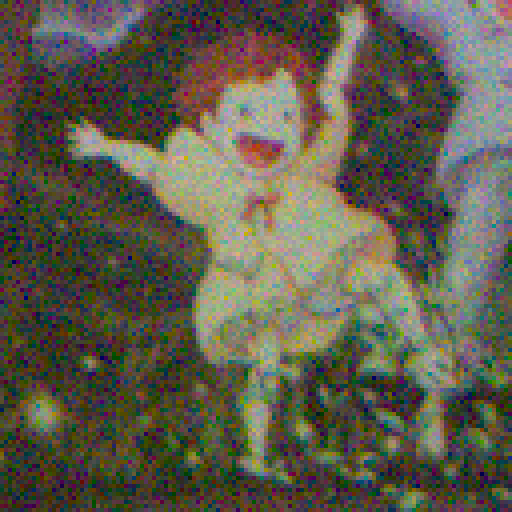} \\
		\rotatebox{90}{\qquad PMRID} &
		\includegraphics[width=.29\linewidth,clip,keepaspectratio]{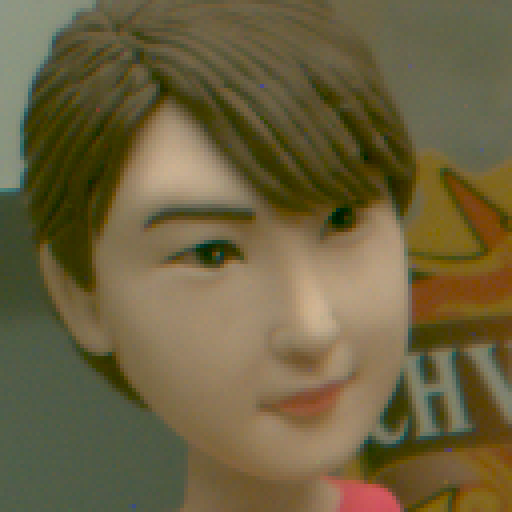} & 
		\includegraphics[width=.29\linewidth,clip,keepaspectratio]{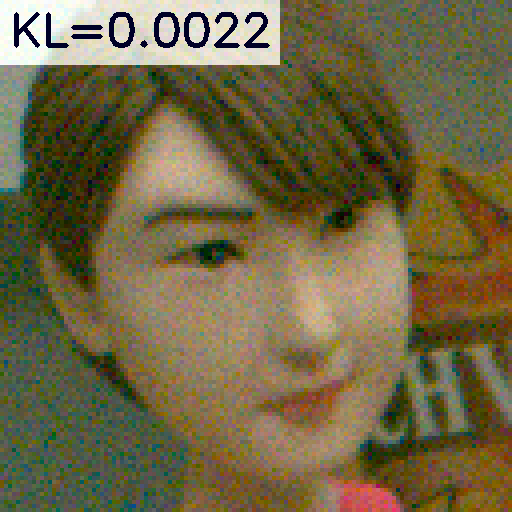} & 
		\includegraphics[width=.29\linewidth,clip,keepaspectratio]{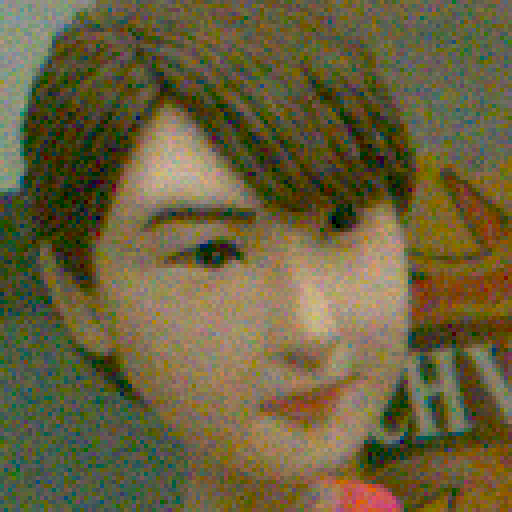} \\
		& Clean & Synthesized & Real Noise \\	
	\end{tabular}
	\vspace{-3mm}
	\caption{Our noise synthesis results on CRVD and PMRID.}
	\label{fig:newdata}
\end{figure}
\noindent\textbf{Noise Model Estimation.} We first evaluate the effectiveness of our contrastive noise estimation model. Our model is trained on synthetic datasets with noise parameters from our well-calibrated camera noise model dataset, and then applied to \textit{unknown} (or not calibrated) sensor estimation. In the training stage, we randomly sample noise parameters $\bm P_i$ from these camera candidates to synthesize noisy images. Our noise estimation model predicts  $(K, \sigma, \mu_c, \sigma_{r})$ and these parameters are supervised by ground truth $\bm P_i$.
We visualize the linear least-square fitting for our noise model estimation in \fref{fig:estimation_plot}. For the top two figures, we presents the estimation on synthetic noisy images, from which we can see that our contrastive noise estimation model can accurately estimate noise parameters. 
The bottom two figures show the estimated model for two mobile sensors of SIDD dataset. Noting that the part of SIDD dataset we use for synthesis purposes consists of $6$ and $4$ ISO levels for Samsung S6 and LG G4, respectively. We observe that the noise parameters estimated by our model apparently form $6$ and $4$ clusters in \fref{fig:estimation_plot}. This phenomenon supports our estimation model for SIDD cameras.

\begin{table*}[t]
	\centering
	\setlength{\tabcolsep}{0.8mm}
	\small
	\begin{threeparttable}
		\vspace{-2mm}
		\caption{Quantitative denoising results of S6 camera on SIDD dataset. Without seeing any data beyond testing noisy images, our noise synthesis pipeline outperforms other generation methods, and even achieves comparable results with paired real data.}
		\begin{tabular}{p{1cm}<{\centering}p{2cm}<{\centering}p{2cm}<{\centering}p{2cm}<{\centering}p{2cm}<{\centering}p{2cm}<{\centering}p{2cm}<{\centering}p{2cm}<{\centering}p{2cm}<{\centering}}
			\toprule
			ISO&Metrices&AWGN&P-G&Noiseflow~\cite{abdelhamed2019noise}&CANGAN~\cite{chang2020learning}&Paired Data&Ours\\
			\midrule
			
			\multirow{2}{*}{$100$} 
			&PSNR & $50.13  $ &$53.80 $&$51.82 $&$52.85 $&${53.94} $&$\bm{54.12} $ \\
			&SSIM &$0.9809 $&$0.9957 $&$0.9941  $&$0.9947 $&${0.9962} $&$\bm{0.9962} $ \\
			\multirow{2}{*}{$800$} 
			&PSNR&$46.45 $&$48.41 $&$42.75 $&$48.20 $&${48.68} $&$\bm{48.82} $ \\
			&SSIM&$0.9700 $&$0.9935 $&$0.9693 $&$0.9917 $&$\bm{0.9942} $&${0.9941} $ \\
			\multirow{2}{*}{$1600$} 
			&PSNR&$47.29 $&$48.92 $&$41.09 $&$47.93 $&${49.10} $&$\bm{49.11} $ \\
			&SSIM&$0.9638 $&$0.9880 $&$0.9281  $&$0.9853  $&$\bm{0.9889} $&${0.9885} $  \\
			\multirow{2}{*}{$3200$}
			& PSNR&$42.16 $&${43.47} $&$34.85  $&$42.90 $&$\bm{43.61} $&$43.05  $\\
			&SSIM&$0.9429 $&${0.9644} $&$0.8054  $&$0.9621 $&$\bm{0.9653} $&$0.9581 $ \\
			\midrule
			\multirow{2}{*}{All} 
			&PSNR&$47.55 $&$49.91 $&$44.96  $&$49.19 $&${50.10} $&$\bm{50.13}  $\\
			&SSIM&$0.9698 $&$0.9896 $&$0.9517 $&$0.9879 $&$\bm{0.9902} $&${0.9891} $ \\
			\bottomrule
		\end{tabular}
		\label{tab:realdenosing}
	\end{threeparttable}

\end{table*}

\begin{figure*}[h!] \small
	\centering
	\setlength{\tabcolsep}{1pt}
	\vspace{-2mm}
	\begin{tabular}{cccccccc}
		Input & AWGN & P-G & Noiseflow~\cite{abdelhamed2019noise} & CANGAN~\cite{chang2020learning} & Paired Data & Ours & GT\\
		\multicolumn{1}{m{.112\linewidth}}{\includegraphics[width=\linewidth,clip,keepaspectratio]{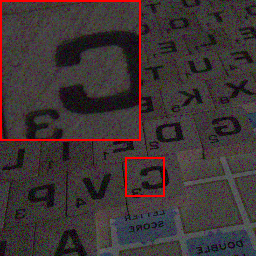}} &
		\multicolumn{1}{m{.112\linewidth}}{\includegraphics[width=\linewidth,clip,keepaspectratio]{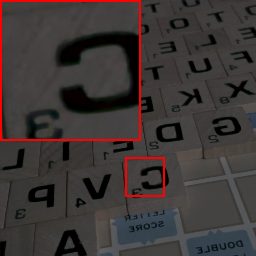}} & \multicolumn{1}{m{.112\linewidth}}{\includegraphics[width=\linewidth,clip,keepaspectratio]{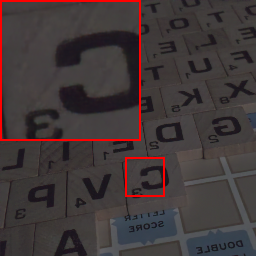}} & \multicolumn{1}{m{.112\linewidth}}{\includegraphics[width=\linewidth,clip,keepaspectratio]{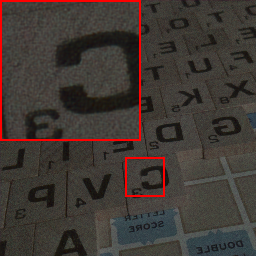}} & \multicolumn{1}{m{.112\linewidth}}{\includegraphics[width=\linewidth,clip,keepaspectratio]{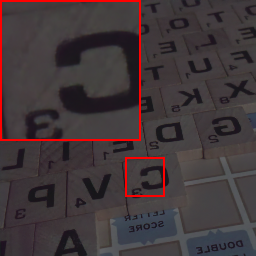}} & \multicolumn{1}{m{.112\linewidth}}{\includegraphics[width=\linewidth,clip,keepaspectratio]{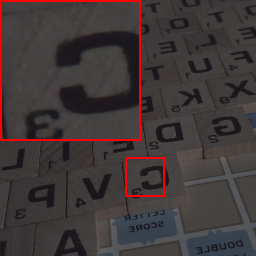}} & \multicolumn{1}{m{.112\linewidth}}{\includegraphics[width=\linewidth,clip,keepaspectratio]{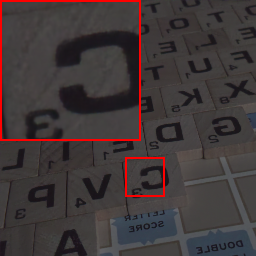}}&
		\multicolumn{1}{m{.112\linewidth}}{\includegraphics[width=\linewidth,clip,keepaspectratio]{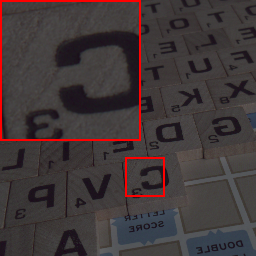}}\\
		\multicolumn{1}{m{.112\linewidth}}{\includegraphics[width=\linewidth,clip,keepaspectratio]{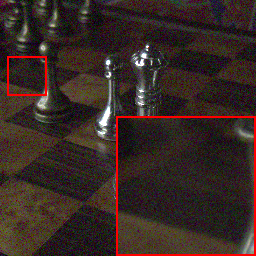}} &
		\multicolumn{1}{m{.112\linewidth}}{\includegraphics[width=\linewidth,clip,keepaspectratio]{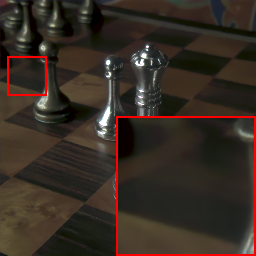}} & \multicolumn{1}{m{.112\linewidth}}{\includegraphics[width=\linewidth,clip,keepaspectratio]{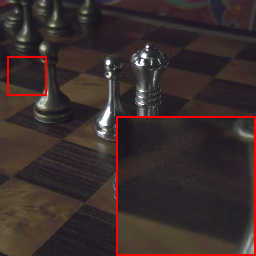}} & \multicolumn{1}{m{.112\linewidth}}{\includegraphics[width=\linewidth,clip,keepaspectratio]{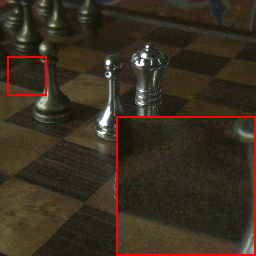}} & \multicolumn{1}{m{.112\linewidth}}{\includegraphics[width=\linewidth,clip,keepaspectratio]{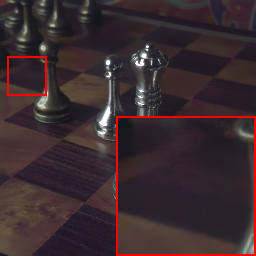}} & \multicolumn{1}{m{.112\linewidth}}{\includegraphics[width=\linewidth,clip,keepaspectratio]{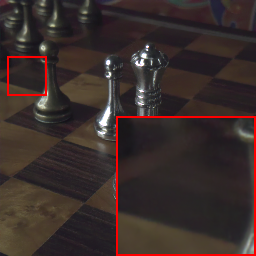}} & \multicolumn{1}{m{.112\linewidth}}{\includegraphics[width=\linewidth,clip,keepaspectratio]{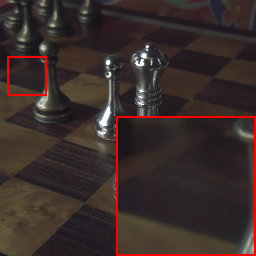}}&
		\multicolumn{1}{m{.112\linewidth}}{\includegraphics[width=\linewidth,clip,keepaspectratio]{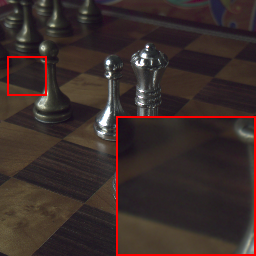}}\\

	\end{tabular}
	\vspace{-2mm}
	\caption{The SIDD~\cite{abdelhamed2018high} S6 denoising results for AWGN/P-G/Noiseflow/CANGAN/Paired Data/Ours are shown from left to right.}
	\label{fig:realdenosing}
	\vspace{-2mm}
\end{figure*}

\vspace{1mm}
\noindent\textbf{Noise Synthesis on SIDD.}
To evaluate our pipeline on noisy image synthesis, we compare it with several state-of-the-art noise modeling methods, including: 1) AWGN noise model 2) P-G noise model, 3) Noiseflow~\cite{abdelhamed2019noise} and 4) CANGAN~\cite{chang2020learning}. Among these methods, AWGN and P-G are commonly used statistical noise models, Noiseflow is a normalizing flow based noise modeling methods, and CANGAN is a representative GAN based noise generation model. The training of Noiseflow and CANGAN requires noisy/clean image pairs. We test all methods in SIDD dataset under different ISOs, and synthesize noise patches of $4\times64\times64$. The noise synthesis accuracy of all compared methods and our pipeline are listed in \tref{tab:synthesis}. By comparing all the methods, it can be seen that our generation pipeline provides promising performance, even if we have never seen any data beyond testing sets. This is partially due to the accurate contrastive noise estimation, and partially by a more realistic fine-grained noise model which carefully considers the image formation process. Though CANGAN also achieves good performances, it requires paired training data and an inference noisy image which has the same setting with the targeted one.
 \fref{fig:synthesis} shows the visualization of synthetic noisy images for all compared noise models and our method. It implies that our pipeline generates more visually realistic noise patches.

\vspace{1mm}
\noindent\textbf{Noise Synthesis on CRVD and PMRID.} We also provide synthesized noisy images on other datasets. Given an noisy image and its corresponding clean one, we first estimate the noise parameters by feeding our model another noisy image which has the same ISO with the targeted image. Then, we use the estimated noise parameters to generate noise on the clean image. As shown in \fref{fig:newdata}, our model produces realistic noise. Please note that none of the cameras used in CRVD, PMRID and SIDD are included in our training data, which means the results can verify the generalization of our pipeline.

\subsection{Applications on Real Image Denoising}
Here, we use the noise synthesis methods (AWGN, P-G, Noiseflow, CANGAN and ours) described in Section~\ref{sec:estimation} to generate synthetic training datasets. Then, these datasets are used to train a common denoising UNet~\cite{ronneberger2015u}, aiming to evaluate the superiority of our model estimation and noise generation pipeline in downstream denoising application. Besides training on synthetic data, we also perform denoising experiments trained on real paired dataset.

 Real image denoising experiments are conducted on SIDD S6 Dataset. We directly use the pretrained synthesis model of Noiseflow and CANGAN, and sample $4\times512\times512$ noisy patches for all methods.
Quantitative results are shown in \tref{tab:realdenosing}. It can be inferred that owing to the high quality training data generated by our noise synthesis pipeline, the denoising results of our method surpasses all compared methods in terms of both pixel-wise accuracy and structural similarity. Another observation is that P-G outperforms CANGAN, which is opposite to the  result of noise estimation. The reason is that statistical models including AWGN, P-G and our model can feed the denoiser with a wider range of noise under continuous ISO values. Besides, we would like to stress that though our synthesis pipeline is built solely on noisy SIDD testing data, it is surprising that our model give similar results  compared with paired real data. These results demonstrate the effectiveness of our method. \fref{fig:realdenosing} shows the denoising visualization of all methods, which indicates that our generation pipeline can practically benefit denoising of real photographs.

\subsection{Ablation Study}
In this section, we perform more experiments to verify the effectiveness of our contrastive noise model estimation framework. We claim that the contrastive learning manner helps the model to learn parameters for separable noise components, and the fine-grained noise model also contributes to better noise synthesis. Therefore, we conduct ablation study, by removing the contrastive loss and replacing the fine-grained noise model with the predominant Hetero-G. Denoising experiments are conducted for each case. As indicated in \tref{tab:ablation}, our full model achieves better  results, which further validate the superiority of our contrastive learning strategy and fine-grained noise model.

\begin{table}[t]
	\centering
	\setlength{\tabcolsep}{6mm}
	\small
	\begin{threeparttable}
		\caption{The ablation study on our contrastive loss and fine-grained noise model.}
		\begin{tabular}{ccc}
			\toprule
			Setting & PSNR & SSIM  \\
			\midrule
			w/o $\mathcal L_\text{contrastive}$&$49.03$&$0.9868$\\
			Hetero-G &${50.04}$&${0.9874}$ \\
			Ours &$\bm{50.13}$&$\bm{0.9891}$ \\
			\bottomrule
		\end{tabular}
		\label{tab:ablation}
	\end{threeparttable}
\end{table}

\section{Conclusion}
\label{sec:conc}
In this paper, we propose a novel noise synthesis pipeline by estimating camera-specific noise models with only testing data. Our method is based on a fine-grained physics-based noise model, and we design a  noise estimation model which is learned in a contrastive manner. Without seeing any paired images or calibration data, our pipeline can achieve competitive results with state-of-the-art noise synthesis methods. It is inspring that given only testing noisy images, our model estimation and noise synthesis pipeline can be directly used in the modeling of other unknown cameras without retraining. Our model is potential to facilitate other applications, including low-light enhancement, which will be remained as our future work.

\vspace{-1.5mm}
\section{Limitation Discussion and Broader Impact}
\vspace{-1mm}
Our model estimation and noise synthesis pipeline aims at estimating noise models of unknown sensors. However, our current model is only used for bayer CFA, and have not extended to non-bayer CFAs like X-Trans. Thus it would be risky if we are not sure about the sensor CFA. Our work has no broader impact.

\vspace{1mm}
\noindent\textbf{Acknowledgments}
This work was supported by the National Natural Science Foundation of China under Grants No.~62171038, No.~61827901, and No.~62088101.

{\small
\bibliographystyle{ieee_fullname}
\bibliography{egbib}
}

\end{document}